\documentclass[11pt,a4paper]{article}
\usepackage[]{geometry}
\usepackage{amsmath,amsthm,amssymb}
\usepackage{graphicx,tabu}
\usepackage[pagebackref=false]{hyperref}
\usepackage[active]{srcltx}
\usepackage[affil-it]{authblk}
\usepackage[titles]{tocloft}

\numberwithin{equation}{section}
\newcommand{\be}{\begin{equation}}
\newcommand{\ee}{\end{equation}}
\newcommand{\bea}{\begin{eqnarray}}
\newcommand{\eea}{\end{eqnarray}}
\newcommand{\nn}{\nonumber}
\def\f{f_\infty}
\def \L {\tilde L}

\begin{document}
\begin{titlepage}
{\title{\bf{Holographic entanglement entropy from  minimal surfaces with/without extrinsic curvature}}}
\vspace{.5cm}
\author{Ahmad Ghodsi \thanks{a-ghodsi@ferdowsi.um.ac.ir}}
\author{Mohammad Moghadassi \thanks{m.moghadassi@stu-mail.um.ac.ir} }
\vspace{.5cm}
\affil{ Department of Physics, Ferdowsi University of Mashhad,    
\hspace{5.5cm} P.O.Box 1436, Mashhad, Iran}
\renewcommand\Authands{ and }
\maketitle
\vspace{-12cm}
\begin{flushright}
\end{flushright}
\vspace{10cm}
\begin{abstract}
In this paper we show that in addition to the known minimal surfaces which appear in the literature for computing the holographic  entanglement entropy, there are other minimal surfaces with non-zero extrinsic curvature.
We use the approach of regularization procedure presented by Fursaev et al in \cite{Fursaev:2013fta}, to compute the quadratic and cubic  curvature invariants on manifolds with squashed cones. The results can be used  to find the leading and universal terms of the holographic entanglement entropy to understand which solution corresponds to the actual minimal entropy.
\end{abstract}
\thispagestyle{empty}
\end{titlepage}
\setcounter{footnote}{0}
\addtocontents{toc}{\protect\setcounter{tocdepth}{4}}
\setcounter{secnumdepth}{4}
\noindent\hrulefill
\tableofcontents
\noindent\hrulefill
\vspace{1.0cm}
\section{Introduction}
The entanglement entropy of a  sub-system in a quantum field theory, is a non-local quantity which measures how this sub-system is correlated with its complement. A holographic description of entanglement entropy first proposed in \cite{Ryu:2006bv} and \cite{Ryu:2006ef}. They argue that the leading divergence of the entanglement entropy of a region $A$ in a $CFT_{d+1}$ is proportional to the area of a $d$-dimensional minimal surface in $AdS_{d+2}$ whose boundary is given by $\partial A$. Generally one can find a series solution in terms of $\ell/\epsilon$ for entanglement entropy, where $\ell$ is the typical length of $\partial A$ and $\epsilon$ is the UV cut-off. In odd dimension $d$, there is a logarithmic divergence $\ln{\ell/\epsilon}$, known as the universal term, which its coefficient is proportional to the central charges of dual CFT.

Usually in computing the entanglement entropy in a quantum field theory one uses the replica trick by introducing a conical singularity. In the context of AdS-CFT one may use a similar  trick by deforming the bulk metric so that the minimal surface contains a conical singularity.
One of the recent works to derive the holographic prescription of \cite{Ryu:2006bv} is the proposal of  \cite{Lewkowycz:2013nqa}. They show that in Einstein gravity the minimal surfaces can be found from the bulk equations of motion when we demand no singularity.

To extend this prescription for higher derivative gravities several attempts have been done.
For example in \cite{Bhattacharyya:2013jma} it has been shown that the method used in \cite{Lewkowycz:2013nqa} only works when the extrinsic curvature is small in Gauss-Bonnet gravity and results coincide with the known Jacobson-Myers entropy \cite{Jacobson:1993xs}, which has established already as a correct entanglement entropy formula \cite{Hung:2011xb}.

The other approach presented in \cite{Fursaev:2013fta}, proposes  a regularization procedure which has been developed to compute the integral curvature invariants on manifolds with squashed cones. By using this method one finds a holographic formula for entanglement entropy when the bulk gravity includes quadratic curvature terms. In special case of Gauss-Bonnet gravity this approach again gives the correct Jacobson-Myers entropy result. For further developments of this method to other theories of gravity see for example \cite{Bhattacharyya:2013gra, Dong:2013qoa, Camps:2013zua, Alishahiha:2013zta, Erdmenger:2014tba, Bhattacharyya:2014yga, Bueno:2014oua, Miao:2014nxa, Guo:2015uqa, Azeyanagi:2015uoa}.

In this paper we have used the approach of \cite{Fursaev:2013fta} to find a formula for entanglement entropy of the  six derivative gravity theories.  Using our results and the known formula for the four derivative gravity theories we will be able to study the holographic entanglement entropy of cylindrical and spherical surfaces which are living on the dual conformal field theories. We also compute the leading and the universal terms corresponding to the extremal surfaces with/without  extrinsic curvature.
We will find two different types of solutions for these extremal surfaces, which exist in different regions of the parameter space of each theory. These regions can be found  by demanding of positivity of the leading order terms in the entropy and reality condition of solutions. We will argue about how we can choose which solution gives the correct result i.e. the minimal entanglement entropy.

Since we have used and developed the method of regularization procedure of Fursaev et al in \cite{Fursaev:2013fta} so let us review very briefly the main results of it:

1. Consider a geometry $\mathcal{M}$ and construct an orbifold $\mathcal{M}_n$ made by cutting $\mathcal{M}$ along a co-dimension one hyper-surface in  $\mathcal{M}$, then glue $n$ identical copies (replica method). The new geometry $\mathcal{M}_n$ has a conical singularity located at $r=0$  with the following metric (according to the replica method we need to set a period of $2\pi n$ for angular coordinate $\tau$)
\be\label{geo1}
ds^2=r^2d\tau^2+dr^2+ \gamma_{ij}(r,\tau;x) dx^idx^j\,,
\ee
where ${x^i}$'s are the coordinates of the co-dimension two surface $\Sigma$ and $\gamma_{ij}$ is its intrinsic metric. This surface is chosen according to the entangling surface which we are interested to study in the dual conformal field theory. It can have different topologies in different dimensions. In this paper we have considered the cylinder and $n$-Sphere  geometries embedded inside the $AdS$ space-time.

2. The geometry (\ref{geo1}) includes the curvature singularity as well as the conical one. So in \cite{Fursaev:2013fta} a regularized metric $(\tilde{\mathcal{M}}_n)$ has been replaced by the above geometry as
\be\label{geo2}
ds^2= r^2 d\tau^2 + \frac{r^2+b^2n^2}{r^2+b^2} dr^2+(a+r^{n}c^{1-n}\cos\tau)^2 ds^2_{\Sigma}\,.
\ee

3. Inserting the regularized geometry into the integral curvature invariants gives rise to a result with an asymptotic series expansion in terms of parameter $b$ when $b\rightarrow 0$. Sending $n\rightarrow 1$ the result of each integral may contain a term which is proportional to $b^0 (1-n)$ (i.e. independent of the regularization parameter). This limiting procedure in \cite{Fursaev:2013fta} suggests that one may perform the following replacement
\bea\label{intr}
\int_{\tilde{\mathcal{M}}_n} d\tau dr d^{d-2}x \sqrt{g} R \rightarrow n \int_{\mathcal{M}} d\tau dr d^{d-2}x \sqrt{\mathcal G} \mathcal R + 4\pi (1-n) \int_{\Sigma} d^{d-2}x {\sqrt{\gamma}} + \mathcal{O}((1-n)^2)\,.
\eea
The left hand side describes the value of a curvature invariant term (here the Einstein-Hilbert term or in general any action constructed out of curvature invariants) on a regularized geometry $\tilde{\mathcal{M}}_n$, i.e. $I[(\tilde{\mathcal{M}}_n)]$. The regularized geometry is given in equation (\ref{geo2}) with periodicity $\tau\sim\tau+2\pi n$. The right hand side contains two main terms. The first term $I[\mathcal{M}]$ is equal to the same action as we begin with but computed on the non-regularized manifold $\mathcal{M}$ with metric given by (\ref{geo1}) and again $\tau\sim\tau+2\pi n$. The second term is a contribution coming from the near region of the tip of the cone in the singular manifold $\mathcal{M}$. This last term in above equation is noting but the area of the co-dimension two surface times the deficit angle of the cone. Now the holographic entanglement entropy can be computed by $S_{HEE}=n\partial_n(I[(\tilde{\mathcal{M}}_n)]-I[\mathcal{M}])_{n\rightarrow 1}$. This explains us why we need to expand $I[(\tilde{\mathcal{M}}_n)]$ up to $O(n-1)$. Explicitly starting from (\ref{intr}), the holographic entanglement entropy will be proportional to the area of the co-dimension two surface which ends on the entangling region as conjectured by Ryu and Takayanagi.
The same computation has been done in \cite{Fursaev:2013fta} for quadratic curvature terms 
\bea\label{intr2}
\int_{\tilde{\mathcal{M}}_n} d\tau dr d^{d-2}x \sqrt{g} R^2 \!\!\!\!&\rightarrow &\!\!\!\! n \int_{\mathcal{M}} d\tau dr d^{d-2}x \sqrt{\mathcal G} \mathcal R^2 \nn\\
\!\!\!\!&+&\!\!\!\! 8\pi (1-n) \int_{\Sigma} d^{d-2}x {\sqrt{\gamma}} \mathcal R + \cdots\,,\nn \\
 \int_{\tilde{\mathcal{M}}_n} d\tau dr d^{d-2}x \sqrt{g} R_{\mu\nu}^2 \!\!\!\!&\rightarrow &\!\!\!\! n \int_{\mathcal{M}} d\tau dr d^{d-2}x \sqrt{\mathcal G} \mathcal R_{\mu\nu}^2 \nn\\
\!\!\!\!&+&\!\!\!\! 4\pi (1-n) \int_{\Sigma} d^{d-2}x {\sqrt{\gamma}} (\mathcal R_{\mu\nu}n^\mu_i n^{\nu}_i-\frac12 K_i^2) \!+\! \cdots\,,\nn \\
 \int_{\tilde{\mathcal{M}}_n} d\tau dr d^{d-2}x \sqrt{g} R_{\mu\nu\alpha\beta}^2 \!\!\!\!&\rightarrow &\!\!\!\! n \int_{\mathcal{M}} d\tau dr d^{d-2}x \sqrt{\mathcal G} \mathcal R_{\mu\nu\alpha\beta}^2\nn \\ 
\!\!\!\!&+&\!\!\!\! 8\pi (1-n) \!\!\int_{\Sigma}\!\! d^{d-2}x {\sqrt{\gamma}} (\mathcal R_{\mu\nu\alpha\beta}n^{\mu}_i n^{\nu}_j n^{\alpha}_i n^{\beta}_j-{K^i_{\mu\nu}} {K_i^{\mu\nu}}) + \cdots,
\eea
where $n^{\mu}_{i}$'s are the unit mutually orthogonal normal vectors to $\Sigma$ ($i=1,2$ for  a co-dimension two surface) and  $K^i_{\mu\nu}$ are components  of extrinsic curvature tensor defined as
\be\label{exc}
K_{\mu\nu}^{i}=h^{\alpha}_{\mu} h^{\beta}_{\nu} \nabla_{\beta}n^{i}_{ \alpha} \,,\qquad\qquad
h^{\alpha}_{\mu}=\delta^{\alpha}_{\mu}-n^{i}_{ \mu}n_{i}^{\alpha}\,.
\ee

4. To find the holographic entanglement entropy $S(\Sigma)$ for a surface $\Sigma$ in a quadratic curvature bulk gravity
\be
I[\mathcal{M}]=-\int_{\mathcal{M}} d^{d}x\sqrt{g}\Big(\frac{R}{\kappa}+2\Lambda+ a R^2+b R_{\mu\nu}^2+c R_{\mu\nu\alpha\beta}^2\Big)\,,
\ee 
we first compute the partition function. In semiclassical approximation the partition function is related to the action by $-\ln Z = I[\mathcal{M}]$. In replica method we need to compute the value of $I[\tilde{\mathcal{M}}_n]$ by prescription mentioned above and then use the following relation
\be
S_{HEE}(\Sigma)=(n\partial_{n}-1) I[\tilde{\mathcal{M}}_n]_{n\rightarrow 1}\,.
\ee
Here the conical singularity of $\tilde{\mathcal{M}}_n$ is located on a minimal co-dimension two hypersurface $\tilde{\Sigma}$ which ends on entangling surface $\Sigma$ on the boundary.
Finally one gets the following formula for entanglement entropy
\be \label{SEE}
S_{HEE}(\Sigma)=4\pi\int_{\tilde{\Sigma}}d^{d-2}x\sqrt{\gamma}\Big(\frac{1}{\kappa}+ 2a\mathcal R + b\mathcal R_{\mu\nu}n^\mu_i n^{\nu}_i+2c\mathcal R_{\mu\nu\alpha\beta}n^{\mu}_i n^{\nu}_j n^{\alpha}_i n^{\beta}_j -\frac{b}{2}K_{i}^2-2c K^{i}_{\mu\nu}K_{i}^{\mu\nu}\Big),
\ee
where the four first terms are nothing but the Wald's entropy and the last two terms are corrections due to existence of a non-vanishing extrinsic curvature of the minimal hyper-surface.

The organization of the paper is as follows: In section 2 we review the holographic entanglement entropy in new massive gravity and find the differential equation which gives the minimal surface. We find two solutions for this equation with/without extrinsic curvature and compute the related leading terms and universal terms. In section 3 we add extensions to new massive gravity to study the behavior of the entanglement entropy in the presence of the new curvature cubed terms. To compute the integrals of curvature invariants we use the method reviewed above. In section 4 we revisit the general curvature square terms and in section 5 we do an exercise with quasi-topological gravity in five dimensions. In each section we discuss about the solutions in terms of their entanglement entropy in different regions of the parameter space. In last section we summarize and discuss about the results.

\section{New Massive Gravity (revisited)}
In \cite{Bhattacharyya:2013gra} by using the method of squashed cones,  the entanglement entropy in New Massive Gravity (NMG) \cite{Bergshoeff:2009hq} has been studied (see also \cite{Erdmenger:2014tba}). They show that by using a bulk metric constructed using the Fefferman-Graham expansion, there exist a minimal surface  and its corresponding entanglement entropy can be computed just by using the Wald's entropy. Here in this section we show that in addition to the known  minimal surface, there is another minimal surface with non-zero extrinsic curvature and we use the squashed cone method to compute the entanglement entropy.

\begin{figure}[ht]
\centering
\includegraphics[bb=0 0 150 150 , scale=1.2]{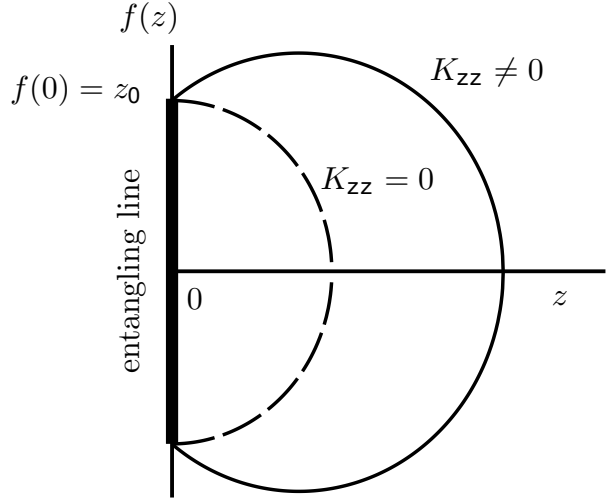}
\caption{The dashed curve corresponds to $f_1(z)$ and the solid one is $f_2(z)$ with non-zero extrinsic curvature.}
\label{fig1}
\end{figure}

In three dimensions we consider a line as an entangling region, which lies on the boundaries of $AdS_3$ space-time located at $z=0$. We Start from the $AdS_3$ metric 
\be \label{ads3}
ds^2=g_{\mu\nu}dx^\mu dx^\nu=\frac{\tilde L^2}{ z^2} \left( dt^2+dr^2+dz^2 \right)\,,
\ee
where $\tilde L$ is the $AdS_3$ radius. 
The Euclidean version of the NMG action is written as \cite{Sinha:2010ai}
\begin{align}
S=-\frac{1}{2\ell_p} \int d^3 x \sqrt{g} \left[ R+\frac{2}{L^2}+4\lambda L^2
\left( R_{\mu\nu} R^{\mu\nu} -\frac38 R^2 \right) \right]\,,
\end{align}
where $L=\tilde L\sqrt{\f}$  and $\f$ satisfies $1-\f+ \f^2 \lambda=0$, using the fact that (\ref{ads3}) is a solution for the NMG equations of motion. 
To compute the entanglement entropy we use (\ref{SEE}), where for NMG it reduces to 
\bea
S^{NMG}_{EE}=\frac{2\pi}{\ell_p} \int dz \sqrt{\gamma} \left[ 1+4\lambda L^2 \Big(\mathcal R_{\mu\nu} n_i^\mu n_i^\nu
-\frac34\mathcal R -\frac12 K_{i}^2 \Big) \right]\,.
\label{seenmg}
\eea
To find the geometry of the extremal surface we must take $t=0$ and $r=f(z)$, then extremize (\ref{seenmg}) to find a suitable value for $f(z)$. The strategy we will use through our computations is finding the tangent vectors at first and then making orthogonal unit vectors. In three dimension there is only one tangent vector which is given by $e_z^a=\partial{x^a}/\partial{z}=( 0,1,f'(z))$.
Then we can build the induced metric of the minimal surface, $\gamma_{\mu\nu}$, and also its orthonormal normal vectors $n_{i\mu}$
\bea
&&ds_\gamma^2=\gamma_{\mu\nu} dx^\mu dx^\nu=\frac {{\tilde L}^{2}   \left(f'(z) ^{2}+1 \right)}{z^2}dz^2\,, \nn \\
&&n_{1\mu}=\Big[ 0,-\frac { \tilde L f'(z) }{z\sqrt { f'(z)  ^{2}+1}},\frac {\tilde L}{z\sqrt {f'(z) ^{2}+1}} \Big]\,,\qquad
 n_{2\mu}= \Big[ \frac{\L}{z},0,0 \Big]\,.
 \eea
 For each normal vector  we have a corresponding extrinsic curvature. For example its $zz$ component is given by (the other non-zero components are proportional to this component when using equation (\ref{exc}))
\begin{align}
 K_{zz}^1=\frac {\L \left(  f'(z) ^{3}- zf''(z) +f'(z) \right)}
{{z}^{2} (f'(z) ^{2}+1)^\frac52}\,,\qquad\qquad
K_{zz}^2=0\,.
\end{align}
The vanishing of $K^2_{zz}$ is a consequence of having a Killing vector in the time direction. This is happening in all computations in this paper so we will drop the index $i$ for simplicity in the up-comping computations. By inserting the above results into \eqref{seenmg}, 
and after extremizing we find the following non-linear differential equation
\begin{align}
&\f \lambda \big((10 f''^3 -60 f'^2 f''^3 +40 f''' f'^3 f'' +40 f''' f' f'' -4 f^{(4)} f'^4 -4 f^{(4)} -8 f^{(4)}  f'^2) z^3\notag\\&
 +(30 f'^3f''^2+30 f' f''^2-8  f''' -8 f''' f'^4   -16 f''' f'^2 )z^2+(2 -4f'^6  -6 f'^4)f'' z  \notag\\&
 -2 f'^7 -6 f'^5 -6 f'^3 -2  f'\big)
-\big(f'^6  +3 f'^4  +3f'^2  +1 \big)f''z+f'^9+4 f'^7 
+6 f'^5+4 f'^3+f'=0\,.
\end{align}
By a little investigation one can find two possible solutions for this differential equation
\bea
f_1(z)=\sqrt{z_0^2-z^2}\,;\qquad \qquad f_2(z)=\sqrt{z_0^2+2z_0 q z-z^2}\,;\quad q^2=2\lambda \f-1\,,
\eea
where in both cases $z_0=f(z=0)$ is equal to half of the length of the entangling line and the turning points are located at $z_t=z_0$ and $z_t=z_0(q+\sqrt{1+q^2})$ respectively (see figure \ref{fig1}). 
Using the relation $1-\f+\f^2 \lambda=0$ we see that to have a real valued solution we must restrict ourselves to either $\f<0$ or $\f\geq 2$. In special value of $\f=2$ both solutions coincide.
The first solution has been found already in  \cite{Bhattacharyya:2013gra}.

\begin{figure}[ht]
\centering
\includegraphics[scale=.5]{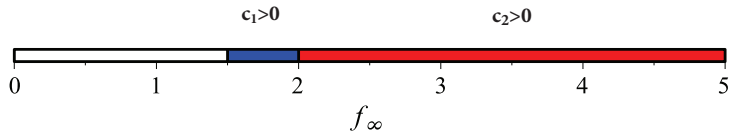}
\caption{The domain of validity (positive value of leading term in entropy) for different extremal surfaces in NMG. The blue domain, $2\geq\f\geq\frac32$,  corresponds to $f_1(z)$ and the red one, $\f\geq 2$,  is for $f_2(z)$.}
\label{NMGfig}
\end{figure}

The main difference between these two solutions is in the value of their extrinsic curvatures 
\be
K^1_{zz} \Big|_{f_1(z)} =0\,,\qquad\qquad
K^1_{zz} \Big| _{f_2(z)} = \frac{\tilde{L} q (z_0^2+2z_0 qz-z^2) }{z^2 z_0^2 (q^2+1)^\frac32}\,.
\ee
Inserting the above results into (\ref{seenmg}) we can compute the entanglement entropy for both cases
\begin{align}
S^{(1)}_{EE}=S_{EE} \Big|_{f_1(z)}&=\frac{2\pi}{\ell_p} \int_{\epsilon}^{z_0} dz\frac{\L z_0}{\sqrt{(z_0^2-z^2) z^2}}(2\lambda \f+1)\notag\\&
=\frac{2\pi}{\ell_p} \frac{L}{\sqrt{\f}}(2\lambda\f+1)\left( \ln (\frac{z_0}{\epsilon})+\frac{1}{4z_0^2}\,z^2 \Big|_{\epsilon}^{z_0}+\cdots \right)\,, \label{furnmg}\\
S^{(2)}_{EE}=S_{EE} \Big|_{f_2(z)}&=\frac{2\pi}{\ell_p} \int_\epsilon^{z_{0}} dz \frac{2 L z_0 \sqrt{2 \lambda}} {z\sqrt{z_0^2+2 q z z_0-z^2}}\notag\\&
=\frac{4\pi}{\ell_p}L \sqrt{2\lambda} \left( \ln (\frac{z_0}{\epsilon}) -\frac{q}{z_0}z \Big|_{\epsilon}^{z_0}+\cdots \right)\,.
 \label{entnmg}
\end{align}
Since the extrinsic curvature vanishes for $f_1(z)$, its corresponding entanglement entropy (\ref{seenmg}) will be equal to the Wald's entropy. 
As we can see for both solutions we  have  universal terms 
\bea \label{cnmg}
&& S^{(1)}_{EE}=\frac{c_1}{3} \ln (\frac{z_0}{\epsilon})\,,\qquad \frac{c_1}{3}=\frac{2\pi}{\ell_p} \frac{L}{\sqrt{\f}}(2\lambda\f+1)=\frac{2\pi L}{\ell_p} \frac{3\f-2}{\f^\frac32}\,,\nn \\
&& S^{(2)}_{EE}=\frac{c_2}{3} \ln (\frac{z_0}{\epsilon})\,,\qquad \frac{c_2}{3}=\frac{4\pi}{\ell_p}L \sqrt{2\lambda}=\frac{2\pi L}{\ell_p} \sqrt{8\frac{\f-1}{\f^2}} \,,
\eea 
where $c_1$ is the central charge of the dual CFT. For $\f\geq 2$ we have a real-valued function for $f_2(z)$ and both central charges are monotonically decreasing functions of $\f$. In this region we have $\frac{2\pi L}{\ell_p} 3\sqrt{2} \geq c_1\geq c_2>0$. Therefore the solution which minimizes the entropy is $f_2(z)$. For $2\geq\f$ only the first solution exists as a minimal surface. In latter case demanding a unitary CFT dual restricts us to $2\geq\f\geq\frac32$, see figure \ref{NMGfig}.
\section{Extended NMG}
To study the properties of entanglement entropy in presence of higher curvature corrections, in this section we consider the Extended NMG (ENMG), first introduced in \cite{Sinha:2010ai}. This is a theory which adds cubic curvature terms into the NMG and the action is given by
\begin{align}
S=-\frac{1}{2\ell_p} \int d^3 x \sqrt{g} \Bigg[ R+\frac{2}{L^2}+4\lambda L^2
\left( R_2 -\frac38 R^2 \right) 
+\frac{17}{12}\mu L^4 \left( R^3- \frac{72}{17} RR_2+\frac{64}{17} R_3\right)\Bigg]\,,
\end{align}
where for simplicity in notation we have defined $R_2=R_{\mu\nu}R^{\mu\nu}$ and $R_{3}=R_{\mu\alpha} R^{\alpha\beta} {R_{\beta}}^{\mu}$. Similar to NMG the $AdS_3$ metric (\ref{ads3})  is a solution of equations of motion for ENMG too. For ENMG we have $ 1-f_\infty +f_\infty^2 \lambda+ f_\infty^3 \mu=0$.

To compute the corrections to the entanglement entropy we need to find the corrections to the equation (\ref{seenmg}) by using the method of squashed cones.
In the presence of the $O(2)$ symmetry i.e. when the extrinsic curvature vanishes,  we have the following relations  for regularization of curvatures \cite{Fursaev:1995ef}
\begin{align}
R^{\mu\nu\alpha\beta}&={\mathcal{R}}^{\mu\nu\alpha\beta}+2\pi(1-n) ( n_{i}^\mu n_i^\alpha n_j^\nu n_j^\beta-n_i^\mu n_i^\beta n_j^\nu n_j^\alpha ) \delta_\Sigma\,,\notag \\
R^{\mu\nu}&={\mathcal{R}}^{\mu\nu}+2\pi(1-n) n_i^\mu n_i^\nu \delta_\Sigma \,,\notag\\
R&=\mathcal R+4\pi (1-n) \delta_\Sigma\,,
\end{align}
where $\mathcal{R}$ curvatures are computed in the regular points $\mathcal M_n/\Sigma$ and $\delta_\Sigma$ is the Dirac delta function. By combining these formulas and using the fact that $\int_{\mathcal M_n/\Sigma} d^dx \cdots=n \int_{\mathcal M} d^dx \cdots $, we find
\bea\label{walds}
\!\!\!\!\!\!\!\!&&\int_{\tilde{\mathcal M}_n} \!\!\!\! d^dx \sqrt{g} R_3=n\int_{\mathcal M} \!\! d^dx \sqrt{\mathcal G}  \mathcal R_3+6\pi (1-n) \int_\Sigma d^{d-2}x \sqrt{\gamma} \mathcal R_{\mu}\,^{\alpha} \mathcal R_{\nu\alpha} n_i^\mu n_i^\nu+ \mathcal{O}\big( (n-1)^2 \big)\,,\nn \\
\!\!\!\!\!\!\!\!&&\int_{\tilde{\mathcal M}_n} \!\!\!\! d^dx \sqrt{g} R^3=n\int_{\mathcal M} \!\! d^dx \sqrt{\mathcal G} \mathcal R^3+12\pi (1-n) \int_\Sigma d^{d-2}x \sqrt{\gamma} \mathcal R^2+ \mathcal{O}\big( (n-1)^2 \big)\,, \\
\!\!\!\!\!\!\!\!&&\int_{\tilde{\mathcal M}_n} \!\!\!\!\! d^dx \sqrt{g} RR_2=n\int_{\mathcal M} \!\! d^dx \sqrt{\mathcal G} \mathcal R \mathcal R_2+4\pi (1-n) \int_\Sigma d^{d-2}x \sqrt{\gamma} \left( \mathcal R_{2}+ \mathcal R \mathcal R_{\mu\nu} n_{i}^\mu n_{i}^\nu \right) + \mathcal{O}\big( (n-1)^2 \big)\,,\nn
\eea
In absence of $O(2)$ symmetry, i.e. when the extrinsic curvature is not zero we need to consider the contribution of extrinsic curvature to the above relations. To do this we use the same steps as \cite{Fursaev:2013fta}. We suppose that the possible independent allowed terms, according to the dimensional analysis, are  those which are constructed either purely from extrinsic curvature or a combination of bulk curvature and extrinsic curvature. These are
\begin{align}\label{pos}
&K^4\,,\;\; K_2^2\,,\;\;K_2K^2\,,\;\;KK_3\,,\;\;K_4\,,\notag\\&
\mathcal R_{\mu\nu}K^{\mu\nu}K\,,\;\; \mathcal R^\nu_\mu K^\rho_\nu K^\mu_\rho\,,\;\;
\mathcal R K_2\,,\;\;\mathcal R K^2\,,\;\;\mathcal R_{\mu\nu\rho\sigma} K^{\mu\rho}K^{\nu\sigma}\,,
\end{align}
where $\mathcal R, \mathcal  R_{\mu\nu}$ and $\mathcal R_{\mu\nu\alpha\beta}$ are defined on minimal co-dimension two hyper-surface $\tilde\Sigma$. For simplicity in use we call the first row as $K$-terms and the second row $RK$-terms. In the case of quadratic curvature terms, the cylindrical and spherical geometries in $d=4$ were enough to compute the coefficients of $K^2$ and $K_2$. As noted in \cite{Fursaev:2013fta}, going to the higher dimensions or considering other geometries such as $S^{d-2}\times R$ do not change the coefficients.

But in the presence of cubic terms due to the enormous number of terms in (\ref{pos}) we need to consider more geometries than before. To compute the contribution of each term we consider cylindrical and spherical geometries inside the $AdS$ space-time as our entangling regions (indeed we used 5 different geometries).
Now let's start from the following regularized metric 
\begin{align}\label{regmet}
ds_{\tilde{\mathcal M}_n}^2=\frac{\tilde L^2}{z^2}\Big(dz^2+r^2d \tau^2+\frac{r^2+b^2n^2}{r^2+b^2} dr^2+ (a+r^n c^{1-n} \cos \tau)^2 ds_{\Sigma}^2\Big)\,,
\end{align}
where $\Sigma$ is the entangling region (cylinder or $n$-sphere).

To compute the entanglement entropy we need to find a similar relation as (\ref{intr2}) for each term in the Lagrangian. For example we start from the following term
\bea\label{RR2gen}
\!\!\!\!\!\!\!\!\!\!\!\!\!\!\!\!&&\int_{\tilde{\mathcal M}_n} d^dx \sqrt{g} RR_2=n\int_{\mathcal M} d^dx \sqrt{\mathcal G} \mathcal R \mathcal R_2+4\pi (1-n) \int_\Sigma d^{d-2}x \sqrt{\gamma} \left( \mathcal R_{2}+ \mathcal R \mathcal R_{\mu\nu} n_{i}^\mu n_{i}^\nu \right) \nn\\
\!\!\!\!\!\!\!\!\!\!\!\!\!\!\!\!&&+\int_\Sigma d^{d-2}x \sqrt{\gamma}\big(A_1 K^4+A_2 K_2 K^2+A_3 K K_3
+A_4 K_2^2+A_5 K_4\big) \nn \\
\!\!\!\!\!\!\!\!\!\!\!\!\!\!\!\!&&+ \int_\Sigma d^{d-2}x \sqrt{\gamma}\big(B_1 \mathcal R K^2\!+\!B_2 K K_{\mu\nu} \mathcal R^{\mu\nu} \!+\! B_3 K_{\mu\nu} \mathcal R^{\nu\alpha} K_{\alpha}^{\mu}\!+\! B_4 \mathcal R K_2\!+\! B_5 K^{\mu\alpha} K^{\nu\beta} \mathcal R_{\mu\alpha\nu\beta}\big)\!,
\eea
where, as we discussed before the second integral on the right hand side of the first line is the Wald contribution. To compute the unknown coefficients, $A_1,\cdots , B_5$, we must compute  each value of integrals in both sides.  Each integral can be written as a series, $\int dz \sum_{n} c_{n} z^n$. Then we can make a set of algebraic equations by looking at the same powers of $z$ on both sides. You can see the final results of coefficients in table 1.
There are some general properties in this table:

1. The values of Wald's entropies have different power expansion other than the value of $K$-terms or $RK$-terms.
So the value of Wald's entropy does not appear in our algebraic equations. Note that it has a non-zero value which equally appears on the integrals of the left hand side.

2. The power expansion of $K$-terms  differs from $RK$-terms too, but the integrals of the left hand side have  both power expansions. Therefore we have two distinct sets of algebraic equations, one for $A_i$'s and one for $B_i$'s.
  
3. There are some simple relations between the values of integrals on the right hand side at each dimension $d$ due to the fact that for n-spheres the Ricci curvature is proportional to the metric. For example in  $AdS_d$, $K\sim d-3$ and $\mathcal R \sim d(d-1)$ therefore $\mathcal R K^2=d KK_{\mu\nu}R^{\mu\nu}$ or $\mathcal R K_2=d K_{\mu\alpha}\mathcal R^{\alpha \nu} K_{\nu}^\mu$ and $\mathcal R K^2-\mathcal R K_2= d (d-1) K^{\mu\nu}K^{\alpha\beta}\mathcal R_{\mu\alpha\nu\beta}$. 

4. With the same reason as mentioned above the value of $KK_3$ and $K_2^2$ are equal independent of dimension for cylinder or $n$-spheres, so we consider just one of them. It is important to remember that this may not be correct for more general entangling regions other than the cylinder or spheres. 

\begin{figure}[ht]
\centering
\includegraphics[scale=.5]{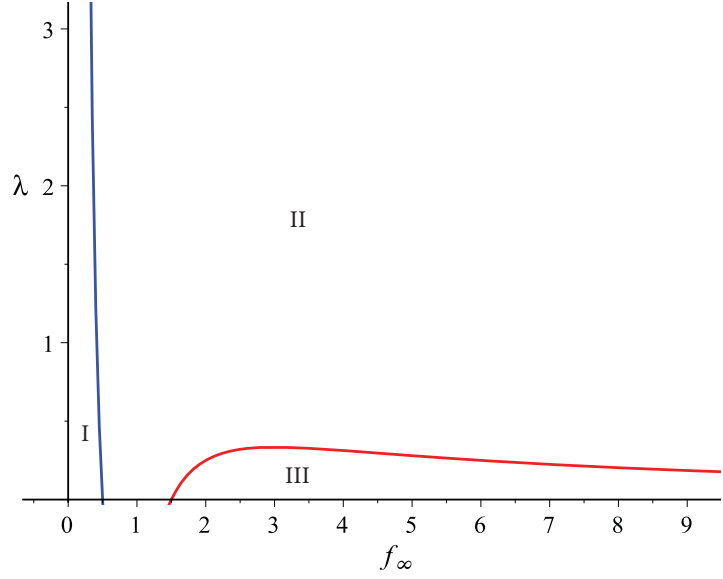}
\caption{To have $c_1>0$, the region I is forbidden for $f_1(z)$. This solution is valid in both regions (II) and (III). The reality condition of $c_2$ restricts $f_2(z)$ to the region (III). In the overlap region III we always have $c_1>c_2$.}
\label{ENMGfig}
\end{figure}

Finally by solving the algebraic equations we find
\begin{align}
&\int_{\tilde{\mathcal M}_n} d^dx \sqrt{g} RR_2= n\int_{\mathcal M} d^dx \sqrt{\mathcal G} \mathcal  R \mathcal R_2\nn\\
&+ 4\pi (1-n) \int_\Sigma d^{d-2}x \sqrt{\gamma} \Big(  \mathcal R_{2}+ \mathcal R \mathcal R_{\mu\nu} n_{i}^\mu n_{i}^\nu-\frac14 K^4+\frac14 K_2K^2 - \frac12 \mathcal R K^2\Big)\,,
\end{align}
Performing the same computations for $R^3$ and $R_3$ terms in the Lagrangian  (see appendix A for a list of integrals) we find 
the final formula for entanglement entropy of ENMG as
\bea
&&S^{ENMG}_{EE}=\frac{2\pi}{\ell_p}
\int_\Sigma\sqrt{\gamma}dz \big[1+4\lambda L^2 \big(\mathcal R_{\mu\nu}n_i^\mu n_i^\nu-\frac12 K^2-\frac34 \mathcal R \big)+\frac{1}{2} \mu L^4\big(3K^4-3K_2K^2 \nn\\
&&+6\mathcal R K^2-16 K K_{\mu\nu}\mathcal R^{\mu\nu}+\frac{17}{2} \mathcal{R}^2 -12 \mathcal{R} \mathcal{R}_{\mu\nu} n^\mu_i n^\nu_i -12 \mathcal{R}_2 +16 \mathcal{R}_{\mu\alpha} \mathcal{R_{\nu}}^{\alpha} n_{i}^{\mu} n_{i}^{\nu} \big)\big]\,.
\eea
Now we can find the minimal surface similar to the NMG case. By extremizing,  we will find two solutions, one with zero and the other with a non-zero extrinsic curvature
\bea
f_1(z)=\sqrt{z_0^2-z^2}\,;\qquad \qquad f_2(z)=\sqrt{z_0^2+2z_0 q z-z^2}\,;\quad q^2=\frac{1-2\lambda \f-3\mu \f^2}{\mu\f^2 -1}\,.
\eea
By knowing these solutions we can compute the universal terms in entanglement entropy
\bea\label{cenmg}
&&S^{(1)}_{EE}=\frac{c_1}{3}\ln\frac{z_0}{\epsilon}\,,\qquad  \frac{c_1}{3}= \frac{2 \pi L(\mu \f^2+2 \lambda \f+1)}{\ell_p\sqrt{\f}}\,,\nn\\
&&S^{(2)}_{EE}=\frac{c_2}{3}\ln\frac{z_0}{\epsilon}\,,\qquad  \frac{c_2}{3}= \frac{4 \pi L}{\ell_p} \sqrt{2(1-\mu \f^2)(\lambda+\mu \f)}\,.
\eea
If we  use the relation $1-\f+\f^2\lambda+\f^3\mu=0$ then we will observe that
to have a real function for $f_2(z)$ and for positive coupling $\lambda$, we must have $\frac13\geq\lambda > 0$ and $\frac{1+\sqrt{1-3 \lambda}}{\lambda}\geq \f \geq \frac{1-\sqrt{1-3 \lambda}}{\lambda}  $. Note that in this interval, $c_2$ is real valued if $\f\geq 1$ which is satisfied automatically. We also have
$c_1^2-c_2^2=\frac{36\pi^2 L^2}{\ell_p^2\f^3}(\lambda \f^2-2\f+3)^2$ so in the region allowed by reality condition, $c_1\geq c_2$ and $f_2(z)$ gives the minimal entropy. For $\lambda>\frac13$ the first solution is the correct one where by demanding a unitary dual CFT we restrict to $\f\geq \frac{\sqrt{1+\lambda}-1}{\lambda}$. The different domains of validity of the solutions has been shown in figure \ref{ENMGfig}.
\section{General curvature square terms}
In previous section we considered a special known action of gravity and its extension in three dimensional space-time and we used a line as our entangling surface. To generalize the study of holographic entanglement entropy to higher dimensions and for more general entangling surfaces such as cylindrical or spherical regions, in this section we consider the general curvature square terms in five dimensions. We start from the following action and use a similar  notation as \cite{Bhattacharyya:2013gra}
\be
S=-\frac{1}{2\ell_p^3}\int d^{5}x\sqrt{g}\Big(R+\frac{12}{L^2}+ \frac{\lambda L^2}{2}(\lambda_1 R_{\mu\nu\alpha\beta}^2+\lambda_2 R_{\mu\nu}^2+\lambda_3 R^2)\Big)\,.
\ee
Let's first consider $S^3$ as an entangling surface, then the non-regularized bulk metric will be
\be
ds^2=\frac{\tilde L^2}{z^2}\Big(d\tau^2+dz^2+d\rho^2+\rho^2(d\theta^2+\sin^2\theta d\phi^2)\Big)\,,
\ee
where $\tilde L= L/\sqrt{\f}$ and here we have $1-\f+\frac13 \lambda \f^2 (\lambda_1+2\lambda_2+10\lambda_3)=0 $ as a constraint on the parameters space.
Doing the same steps as what we did for previous cases, one finds a the differential equation which gives the extremal surface for $S^3$ entangling surface. 
Again there are two solutions  without/with extrinsic curvature
\bea\label{f1f2GB}
f_1(z)=\sqrt{z_0^2-z^2}\,;\qquad  f_2(z)=\sqrt{z_0^2+2z_0 q z-z^2}\,;\quad q^2={\frac {4-(16\lambda_{{1}}+22\lambda_{{2}}+80
\lambda_{{3}})\lambda \f}{-4+5(4\lambda_{{1}}+5\lambda_{{2}}+16\lambda_{
{3}})\lambda \f}}\,.
\eea
Then the leading terms in the entanglement entropy will be
\bea\label{UR2}
\!\!\!\!&&S^{(1)}_{EE}=4a_1(\frac{z_0^2}{\epsilon^2}-\ln\frac{z_0}{\epsilon}+\cdots)\,, a_1={\frac { {\pi }^{2}{L}^{3} \left( 1-2\lambda\f \left( \lambda_{{1}}+2\,\lambda_{{2}}+10\,\lambda_{{3}} \right)  \right)}{\ell_p^3{\f}^{\frac32}}}\,,\\
\!\!\!\!&&S^{(2)}_{EE}=4a_2(\frac{1}{q^2+1}\frac{z_0^2}{\epsilon^2}+\frac{2q}{q^2+1}\frac{z_0}{\epsilon}-\ln\frac{z_0}{\epsilon}+\cdots)\,,  a_2=\frac {{\pi }^{2}{L}^{3}{\lambda}^{\frac32}\left( 4\,\lambda_{{1}}+3\,\lambda_{{2}}
 \right) ^{\frac32}} {2\ell_p^3\sqrt {-4+5\left(4\lambda_{{1}}+5\lambda_{{2}}+16\lambda_{{3
}} \right) \lambda \f}}\nn
.
\eea
For cylindrical geometry with length $H$ we consider
\be
ds^2=\frac{\tilde L^2}{z^2}\Big(d\tau^2+dz^2+d\rho^2+\rho^2 d\theta^2+ du^2\Big)\,.
\ee
The possible solutions are two series solutions as
\bea
f_1(z)=z_0-\frac{z^2}{4z_0}+\cdots\,, \qquad \qquad\qquad f_2(z)=z_0+q z- (1+q^2) \frac{z^2}{4z_0}+\cdots\,,
\eea
with the same value for $q$ as in equation (\ref{f1f2GB}).
Using the above solutions the leading terms of entanglement entropy are
\bea\label{univr2c}
&&S^{(1)}_{EE}=\frac{H}{2 z_0}(4a_1\frac{z_0^2}{\epsilon^2}-c_1\ln\frac{z_0}{\epsilon}+\cdots)\,,\qquad c_1=\frac{\pi^2 L^3}{\ell_p^3\f^\frac32}(1+2\lambda \f (\lambda_1-2\lambda_2-10\lambda_3))\,,\nn\\
&&S^{(2)}_{EE}=-\frac{H}{2 z_0}\frac{\pi^2 L^3\lambda^\frac12}{\ell_p^3\f}(c'_0 \frac{z_0^2}{\epsilon^2}+c'_1 \frac{z_0}{\epsilon}+c'_2\ln\frac{z_0}{\epsilon}+\cdots)\,,\nn\\
&&c'_0=-2 \sqrt{4\lambda_1+3\lambda_2} \sqrt{-4+5(4\lambda_1+5\lambda_2+16\lambda_3)\lambda\f } \,,\nn\\
&&c'_1=-2\sqrt{4\lambda_1+3\lambda_2}\sqrt{4-(16\lambda_1+22\lambda_2+80\lambda_3)\lambda\f}\,,\nn\\
&&c'_2=\frac{8\big((19\lambda_1^2+(31\lambda_2+80\lambda_3)\lambda_1+\frac{141}{16}\lambda_2(\lambda_2+\frac{160}{47}\lambda_3))\lambda\f-\frac{3}{2}\lambda_2-4\lambda_1\big)}{\sqrt{3\lambda_2+4\lambda_1} \sqrt{-4+5 (4\lambda_1+5\lambda_2+16\lambda_3)\lambda\f} }\,.
\eea 

$\bullet{}$ The first solutions $f_1(z)$:
As we see from the results in (\ref{UR2}) and (\ref{univr2c}) the leading terms of both entropies   have the same coefficient $a_1$ independent of the topology of the entangling surfaces. Moreover the coefficients of the logarithmic terms ($a_1$ and $c_1$) exactly coincide with the known results of central charges of the dual CFTs. 
In this case and in the three parameter family of the solutions which specify by $\lambda_{1,2,3}$, we need to have a positive value for $a_1$, which restrict us to $ 1\geq 2\lambda\f ( \lambda_{{1}}+2\,\lambda_{{2}}+10\,\lambda_{{3}} )$.

$\bullet{}$ The second solutions $f_2(z)$: In order to have a real solution in this case we must demand a real value for $q$ which means that $q^2>0$ in equation (\ref{f1f2GB}). Moreover, to have a positive value for leading term of entropy, for spherical region we need to have a real and positive value of $a_2$. In cylindrical case $c'_0$ must be real and negative. 
\subsection{Gauss-Bonnet gravity}
To compare the domain of validity of our two different solutions, it will be easier to choose a specific point in the three parameter family of the solutions. The Gauss-Bonnet gravity is one of the interesting cases (note that it is topological in four dimensions). 

So let's restrict ourselves to $\lambda_1=1, \lambda_2=-4$ and $\lambda_3=1$. 
The constraint explained in the previous section then simplifies to $\lambda=\frac{\f-1}{\f^2}$. For non-zero extrinsic curvature solution and for spherical entangling region, $q=\sqrt{\frac{\f-2}{\f}}$. Two central charges in (\ref{UR2}) simplify to
\be
a_1=\frac{\pi^2 L^3}{\ell_p^3 \f^{\frac52}}(6-5\f)\,,\qquad
a_2={\frac {4\sqrt {2}{\pi }^{2}{L}^{3} ( \f -1) ^{\frac32}}{\ell_p^3{\f}
^{3}}}\,.
\ee
The only consistent solution for $f_2(z)$ happens when $q$ is real-valued. This restricts us to $\f\geq 2$. In this region the leading term of $S^{(2)}_{EE}$ is positive but for $S^{(1)}_{EE}$ it is negative. The only acceptable solution for $f_1(z)$ is in the region where $\frac65\geq\f\geq0$.

In cylindrical geometry from (\ref{univr2c}) and relation $1-\f+\lambda \f^2=0$ we have
\bea
&&a_1=\frac{\pi^2 L^3}{\ell_p^3 \f^{\frac52}}(6-5\f)\,,\qquad c_1=\frac{\pi^2 L^3}{\ell_p^3}\frac{2-\f}{\f^\frac52}\,,\qquad q^2=\frac{\f-2}{\f}\,,\nn \\ 
&&c'_0=\frac{-8\sqrt{2}\pi^2 L^3\sqrt{\f-1}}{\ell_p^3\f^2}\,,\quad
c'_1=qc'_0\,,\quad
c'_2=\frac{-\sqrt{8}\pi^2 L^3\sqrt{\f-1}(\f-2)}{\ell_p^3\f^3 }\,.
\eea
The second solution exists again only for $\f\geq2$ where $S^{(2)}_{EE}$ has a positive leading order coefficient but in this region $S^{(1)}_{EE}$ is negative. For $\frac65\geq\f>0$ only the first solution is allowed. Note that in this region the central charge $c_1$ has a positive value \footnote{For holographic c-theorems specially in odd dimensions see \cite{Myers:2010tj}.}. 

\begin{figure}[ht]
\centering
\includegraphics[scale=.5]{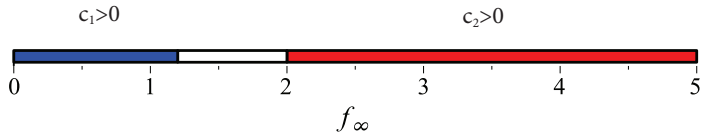}
\caption{The domains of validity for different extremal surfaces in Gauss-Bonnet gravity are two distinct regions. The blue domain for $f_1(z)$ is restricted to $\frac65\geq\f>0$. The red region  for $f_2(z)$ is bounded from below, $\f\geq2$.}
\label{GBfig}
\end{figure}
As we observe from the computations in special case of Gauss-Bonnet gravity, the domain of validity of two solutions is independent of entangling surfaces and specifically it has two distinct regions. Therefore if one demands to have a unitary CFT dual theory, which can be achieved by restricting to the blue region in figure (\ref{GBfig}) then we can ignore the second solution.
\section{Quasi-Topological Gravity}
In three dimensions we studied the extended NMG as a theory with cubic curvature terms. But in this case since the entangling surface was a line, the value of extrinsic curvature was zero when we considered the first solution $f_1(z)$. Note that we can always use the first solution as a check of our calculations because the coefficient of the universal term is a known value of the central charge for the dual CFT. To study a theory with cubic curvature terms and with non-trivial extrinsic curvature we consider the quasi-topological gravity in five dimensions. We will show that while the extrinsic curvature is zero for spherical entangling surfaces it is non-trivial for cylindrical case, therefore quasi-topological gravity can provide us an opportunity to check the results of regularized integrals in appendix A.   
The bulk action is given by \cite{Myers:2010jv}
\bea
&&S=-\frac{1}{2\ell_p^3}\int d^5x \sqrt{g}\left(R+\frac{12}{L^2}+\frac{\lambda L^2}{2} \mathcal L_2+\frac{7\mu L^4}{4} \mathcal{L}_3\right)\,,\nn\\
&&\mathcal L_{2}=R_{\mu\nu\alpha\beta} R^{\mu\nu\alpha\beta}-4 R_{\mu\nu}R^{\mu\nu}+R^2\,,\nn\\
&&\mathcal L_{3}=R^{\mu\nu\rho\sigma} 
R_{\nu\alpha\sigma\beta} R^{\alpha}\!_{\mu}\!^{\beta}\!_{\rho} +\frac38 R_{\mu\nu\alpha\beta} R^{\mu\nu\alpha\beta} R-\frac97 R_{\mu\nu\rho\alpha} R^{\mu\nu\rho}\,_{\beta} R^{\alpha\beta}+\frac{15}{7}R_{\mu\nu\alpha\beta} R^{\mu\alpha} R^{\nu\beta}\nn\\
&&+\frac{18}{7} R_{\mu\alpha} R^{\alpha\beta} R_{\beta}\,^\mu-\frac{33}{14} R_{\mu\nu}R^{\mu\nu} R+\frac{15}{56}R^3\,.
\eea
The entanglement entropy for this case also has been studied in \cite{Bhattacharyya:2013gra} as explained in the section 2, and in \cite{Bhattacharyya:2014yga} using the formula found in \cite{Dong:2013qoa} for quadratic curvature theories.
Both results agree with the universal terms expected for this theory.

Since we have cubic correction terms we need to use the regularized integrals given in appendix A. Considering each term in Lagrangian and replacing its corresponding integrals and after simplification we will find the following relation for holographic entanglement entropy of quasi-topological gravity
\small
\begin{align}
&S^{QT}_{EE}=\frac{2\pi}{\ell_p^3}\int d^{d-2}x \sqrt{\gamma} \bigg( 1+ 
L^2\lambda \big( \mathcal R+\mathcal R_{\mu\nu\rho\sigma} n_i^{\mu} n_i^\rho n_j^{\nu} n_j^\sigma -2 \mathcal R_{\mu\nu} n_i^{\mu} n_i^{\nu}+K^2-K_2\big) \nn\\
&+\frac{3L^4 \mu }{32} \Big( 15 \mathcal{R}^2\!-\!44 \mathcal R_2 \!+\!7 \mathcal R_{\mu\nu\rho\sigma}  \mathcal R^{\mu\nu\rho\sigma} \!-\! 4\big(11 \mathcal{R}\mathcal{R}_{\mu\nu} 
\!+\!3 \mathcal R_{\alpha\sigma\rho\mu}  \mathcal R^{\alpha\sigma\rho}\!_{\nu} 
\!-\!10 \mathcal R_{\sigma\mu\rho\nu}  \mathcal R^{\sigma\rho}\!-\!18 \mathcal{R_\mu}\!^{\rho} \mathcal{R}_{\rho\nu}\big) n_i^\mu n_i^{\nu}
\nn\\&
+8\big(5 \mathcal R_{\mu\rho}  \mathcal R_{\nu\sigma} -6 \mathcal R_{\mu\nu\rho\alpha}  \mathcal R^\alpha\!_\sigma-7\mathcal R^{\alpha}\!_{\mu}\!^{\beta}\!_{\rho} \mathcal R_{\nu\alpha\sigma\beta} +\frac74 \mathcal{R} \mathcal{R}_{\mu\nu\rho\sigma} \big)\big( n_i^{\mu} n_i^\rho 
n_j^{\nu} n_j^\sigma-n_i^{\mu} n_i^\sigma n_j^{\nu} n_j^\rho\big)+2 K_4+25 K_2^2 \nn\\
& -38 K_2K^2\!+\!11 K^4+22 \mathcal R K^2\!-\!104 K K_{\mu\nu} \mathcal R^{\mu\nu}  
\!+\!104 K_{\mu\alpha} \mathcal R^{\alpha \nu} K_{\nu}^{\mu}\!-\! 22 \mathcal R K_2\!+\!8 K^{\mu\alpha} K^{\nu\beta} \mathcal R_{\mu\nu\alpha\beta} \Big)\bigg)\,,
\end{align}
\normalsize
where this result includes the Wald's entropy as well as corrections coming from the existence of non-trivial extrinsic curvature. Similar to GB case in five dimension we can find the value of holographic entanglement entropy for different entangling regions.
 
\begin{figure}[ht]
\centering
\includegraphics[scale=.5]{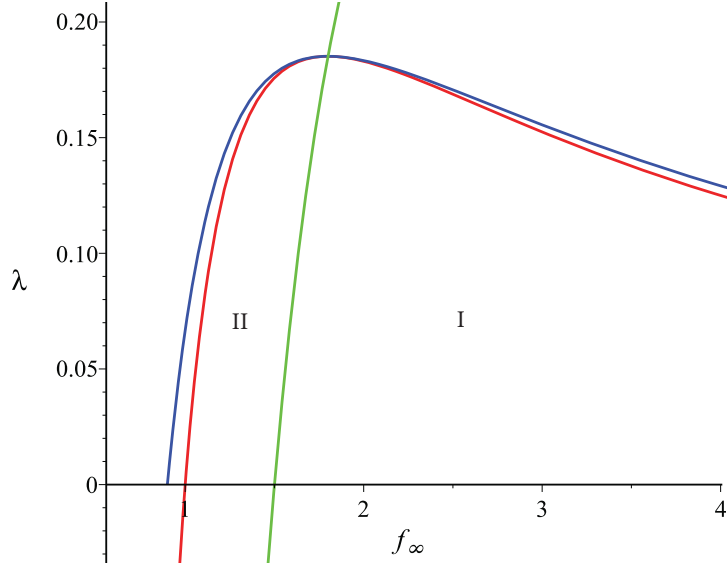}
\caption{The regions allowed by positivity condition of  leading terms  in quasi-topological gravity are the regions (I) and (II). The reality condition of the second solution also restricts us to the region (I).}
\label{QTfig}
\end{figure}

$\bullet$ For spherical region the solutions of the corresponding differential equation for minimal surface are
\bea
f_1(z)=\sqrt{z_0^2-z^2}\,,\qquad\quad f_2(z)=\sqrt{z_0^2+2z_0q z-z^2}\,,\qquad 
q^2=-1+(\lambda\pm\sqrt{\lambda^2+3\mu})\f\,.
\eea
For each solution we find the following entanglement entropies
\begin{align}\label{cqts}
S^{(1)}_{EE}&=\frac{4\pi^2 L^3}{\ell_p^3 \f^{3/2}} (1-6\lambda\f+9\mu\f^2) (\frac{z_0^2}{\epsilon^2}-\ln\frac{z_0}{\epsilon}+\cdots)\,,\\ 
S^{(2)}_{EE}&=\frac{4\pi^2 L^3}{\ell_p^3 \f^{3/2}} (\frac{q^4+2(1-3\lambda\f)q^2+1-6\lambda\f+9\mu \f^2}{(q^2+1)^{1/2}})(\frac{1}{q^2+1}\frac{z_0^2}{\epsilon^2}+\frac{2q}{q^2+1}\frac{z_0}{\epsilon}-\ln\frac{z_0}{\epsilon}+\cdots)\,.\nn
\end{align}
As we see the coefficient of the leading term (and logarithmic term) is the central charge corresponding to the dual conformal field theory of the quasi-topological gravity.
To study the leading order terms one may use the constraint $\mu \f^3+\lambda \f^2-\f+1=0$. To have a positive value for entropies both leading terms must be positive valued. Additionally we demand that the second solution is real valued. The positive value of the leading term in $S^{(1)}$ restricts us to $0\leq\lambda\leq \frac{10\f-9}{15\f^2}\leq\frac{5}{27}$ and therefore $\f\geq\frac{9}{10}$ if we suppose a positive coupling. This corresponds to a region below the blue curve in figure \ref{QTfig}. On the other hand this condition for leading term of $S^{(2)}$ gives $0\leq\lambda\leq\frac{-\f+\sqrt{5\f^2-4\f}}{2\f^2}\leq \frac{5}{27}$ for $\f\geq1$ (the regions (I) and (II) below the red curve in figure \ref{QTfig}). The reality condition for the second solution also restricts us to the region $0\leq\lambda\leq\frac{2\f-3}{\f^2}$ and therefore $\f\geq\frac32$ (the right hand side of the green curve in figure \ref{QTfig}). It is possible to show that in region (I),  the leading term of $S^{(1)}$  is always greater than the leading term of $S^{(2)}$, so the minimal surface corresponds to $f_2(z)$.

$\bullet$ One can also look at the cylindrical entangling region.
The series solutions can be found by solving  the differential equation for $f(z)$ 
\bea
&&f_1(z)=z_0-\frac{z^2}{4z_0}+\cdots\,,\\ \nn 
&&f_2(z)=z_0+q z- (1+q^2) \frac{z^2}{4z_0}+\cdots\,,\qquad 
q^2=-1+(\lambda\pm\sqrt{\lambda^2+3\mu})\f\,.
\eea
The corresponding universal terms of entanglement entropies are
\bea\label{cqtc}
&& S^{(1)}_{EE}=\frac{ H}{2 z_0}(4a_1\frac{z_0^2}{\epsilon^2}-c_1\ln\frac{z_0}{\epsilon}+\cdots)\,, \nn\\
&&c_1= \frac{\pi^2 L^3}{\ell_p^3\f^{3/2}} (1-2\lambda\f-3\mu\f^2) \,,
\qquad a_1= \frac{\pi^2 L^3}{\ell_p^3\f^{3/2}} (1-6\lambda\f+9\mu\f^2) \,,\nn\\
&&S^{(2)}_{EE}=-\frac{2\pi^2 L^3 H}{z_0 \ell_p^3\f^{3/2}}\frac{1}{(q^2+1)^\frac32} (c'_0\frac{z_0^2}{\epsilon^2}+c'_1\frac{z_0}{\epsilon}+c'_2\ln\frac{z_0}{\epsilon}) \,,\nn\\
&&c'_0=q^4+2(1-3\lambda\f)q^2+1-6\lambda\f+9\mu\f^2\,,\qquad c'_1=q c'_0\,,\nn \\
&&c'_2=\frac32 q^6+\frac14 (13-8\lambda \f) q^4 +\frac18 (16-20\lambda \f+51\mu\f^2) q^2 +\frac14 (1-2\lambda \f -3\mu\f^2)\,.
\eea
As we see, by comparing the coefficients of cylindrical region with spherical region, we will find exactly the same conditions as what we found for the spherical entangling surface. We have observed the same behavior when we considered the Gauss-Bonnet gravity in five dimensions. Therefore this suggests that at least for cubic corrections such as quasi-topological gravity (which the equations of motion are quadratic for $AdS$ background) the domain of validity of solutions is independent of topology (here cylindrical and spherical surfaces). We must emphasis that the value of extrinsic curvature is non-zero for cylindrical case when we consider the first solution and this gives us an opportunity to check the regularized integrals in appendix A specially the correction parts including the extrinsic curvature. As we see the value of $c_1$ and $a_1$ exactly agree with known results for central charges found for the dual CFT of quasi-topological gravity.
\section{Conclusion and Summary}

In this paper we have used the procedure for computing the integral curvature invariants on manifolds with squashed cones which was introduced in \cite{Fursaev:2013fta} and we have reviewed it briefly in introduction. This method has been used already to compute the corrections to the Wald's entropy. The corrected terms are constructed out of extrinsic curvature of the minimal surface, and for the Gauss-Bonnet gravity, they coincide with results of Jacobson-Myers entropy. 

Here we apply this method to compute the holographic entanglement entropy for gravities  with cubic curvature terms. Using our results we can compute the leading and the universal terms of entanglement entropy for theories such as extended NMG in three dimensions or quasi-topological gravity in five dimensions. 

In \cite{Fursaev:2013fta} the curvature squared gravities are studied and two different geometries with squashed cone has been used to compute the corrections to the Wald's entropy. These corrections correspond to the existence of extremal surfaces with non-zero extrinsic curvature (\ref{intr2}). 
By going to the higher curvature theories such as the cubic curvature gravities which we have considered in our paper,
one needs to consider corrections of the following form
\be
K^4\,,\;\; K_2^2\,,\;\;K_2K^2\,,\;\;KK_3\,,\;\;K_4\,,\;\;
\mathcal R_{\mu\nu}K^{\mu\nu}K\,,\;\; \mathcal R^\nu_\mu K^\rho_\nu K^\mu_\rho\,,\;\;
\mathcal R K_2\,,\;\;\mathcal R K^2\,,\;\;\mathcal R_{\mu\nu\rho\sigma} K^{\mu\rho}K^{\nu\sigma}\,,\nn
\ee
in addition to the usual Wald's entropy terms. In this regard since we have ten independent new terms (for example see equation \ref{RR2gen}) then we need to consider at least five different geometries with squashed cone. The main difference between our work and \cite{Fursaev:2013fta} is the fact that we have considered the squashed cones inside the $AdS^d$ space-time where $d=5, 6, 7, 8$. Doing this, as we see from the table 1 in the appendix B, the first five terms will have different expansions in term of the radial direction $z$ than the last five terms and therefore with just five geometries we will be able to compute the ten unknown coefficients. 
One must notice that in computing each integral (see appendix A) the integrals over regularized metric have a finite expansion series which must be reproduced by all integrals on the right hand side.
We have computed different possible curvature cubed terms that appear in ENMG or quasi-topological gravity 
and results are summarized in appendix A. We have have also computed the integrals 
for $d=9$ (columns six and twelve in table 1) as a check for our computations. Consequently we believe that the coefficients of integrals in appendix A are universal and independent of the dimension of space-time, at least for cylindrical and spherical entangling surfaces.

In studying the gravity theories including the quadratic or cubic curvature terms, we have observed that in addition to the usual minimal surfaces in the literature for spherical and cylindrical entangling surfaces i.e. 
\be 
f_1(z)=\sqrt{z_0^2-z^2}\,,\qquad f_{1}(z)=z_0-\frac{z^2}{4z_0}+\cdots\,,\nn
\ee 
there are other types of minimal surfaces with non-zero extrinsic curvature (see figure \ref{fig1})
\be
f_2(z)=\sqrt{z_0^2+2z_0 q z-z^2}\,, \qquad f_2(z)=z_0+qz-(1+q^2)\frac{z^2}{4z_0}+\cdots\,.\nn
\ee 
The value of $q$ depends on the theory which we are studying and it restricts us to special regions of parameter space since $f_2(z)$ must be a real function. 
In addition to this restriction, we must take into account a positive value of the leading terms in the entanglement entropy as well, this usually is  equivalent to the existence of a unitary CFT dual when we consider the first type of solutions i.e. $f_1(z)$. We have confirmed this result (value of central charges) everywhere as a check of our computations, see equations (\ref{cnmg}), (\ref{cenmg}), (\ref{UR2}), (\ref{cqts}) and (\ref{cqtc}).

It must be noticed that although we have computed the entanglement entropy for both spherical and cylindrical entangling regions, but the domain of validity of the solutions are exactly the same and is independent of the topology of entangling surface.

As we mentioned, the universal terms in the entanglement entropy will be proportional to the known central charges of the dual CFTs if we consider the first type of the solutions. On the other hand the universal terms of the second type will give rise to other values. Therefore this question arises, which solution corresponds to the correct entropy? or if the first type of the solutions give the correct entropy  then how one can get ride of the second type solutions?

According to the prescription given in \cite{Ryu:2006bv} and \cite{Ryu:2006ef}, the correct entanglement entropy corresponds to an extremal surface which gives the minimum entropy. To understand which solution has the minimal entropy we have compared the leading order terms in the entanglement entropies. 
To summarize our results, let us compare our answers according to the order of curvature terms in the Lagrangians:

$\bullet$ {\bf{Curvature squared Lagrangians:}} 

By looking at the results of NMG (figure \ref{NMGfig}) and GB (figure \ref{GBfig}) we see that the allowed regions of the solutions in the one dimensional parameter space of these theories (imposed by the positivity of the leading terms and reality condition)  are two distinct regions. So if we restrict ourselves to the unitary dual CFTs we can ignore the second type of the solutions.
For these theories, this choice of extremal surfaces, has been imposed by causality constraints in \cite{Erdmenger:2014tba}.

$\bullet$ {\bf{Curvature cubed Lagrangians:}}

The results of ENMG and quasi-topological gravity have been summarized in figures (\ref{ENMGfig}) and (\ref{QTfig}). As we observe, unlike the curvature square Lagrangians, there are regions in the two dimensional parameter space where both solutions can co-exist.
Interestingly, in both theories regardless of entangling surfaces, in this overlap region always the leading term for $S^{(1)}$ is greater than the leading term of $S^{(2)}$ and therefore the minimal entropy corresponds to $f_2(z)$. So in cubic curvature theories we can not ignore the second solution by imposing the unitary constraints.

Consequently, although unitary constraint (unitary dual CFT) may prevent us to deal with the second type of extremal surfaces $f_2(z)$ in curvature squared theories of gravity, but it can not be helpful for cubic curvature Lagrangians. It would be interesting to find a constraint from the  CFT dual side in order to restrict us to the first type of solutions $f_1(z)$, which gives the actual value of entanglement entropy.

\section*{Acknowledgment}
We would like to thank Aninda Sinha and Mohsen Alishahiha for valuable comments on the draft. This work  is supported by Ferdowsi University of Mashhad
under the grant 3/31581 (1393/05/11).
\pagebreak
\appendix
\section{The curvature cubed integrals}
In this appendix we have summarized the value of curvature cubed terms which appear in different Lagrangians
\begin{align}
&\int_{\tilde{\mathcal M}_n} d^dx \sqrt{g} RR_2= n\int_{\mathcal M} d^dx \sqrt{\mathcal G} \mathcal  R \mathcal R_2\nn\\
&+ 4\pi (1-n) \int_\Sigma d^{d-2}x \sqrt{\gamma} \Big(  \mathcal R_{2}+ \mathcal R \mathcal R_{\mu\nu} n_{i}^\mu n_{i}^\nu-\frac14 K^4+\frac14 K_2K^2 - \frac12 \mathcal R K^2\Big)\,,
\end{align}
\begin{align}
&\int_{\tilde{\mathcal M}_n} d^dx \sqrt{g} R^3 = 
n\int_{\mathcal M} d^dx \sqrt{\mathcal G} \mathcal R^3+12\pi (1-n) \int_\Sigma d^{d-2}x \sqrt{\gamma}\, \mathcal R^2 \,,
\end{align}
\begin{align}
&\int_{\tilde{\mathcal M}_n} d^dx \sqrt{g} R_3=
n\int_{\mathcal M} d^dx \sqrt{\mathcal G}  \mathcal R_3+6\pi (1-n) \int_\Sigma d^{d-2}x \sqrt{\gamma} \big(\mathcal R_{\mu}\,^{\alpha} \mathcal R_{\nu\alpha} n_i^\mu n_i^\nu - K K_{\mu\nu} \mathcal R^{\mu\nu} \big)\,,
\end{align}
\begin{align}
&\int_{\tilde{\mathcal M}_n} d^dx \sqrt{g} R^{\mu\nu\rho\sigma} R_{\nu\alpha\sigma\beta} R^{\alpha}\!_{\mu}\!^{\beta}\!_{\rho}= n \int_{\mathcal M} d^dx \sqrt{\mathcal G} \mathcal R^{\mu\nu\rho\sigma} \mathcal R_{\nu\alpha\sigma\beta} \mathcal R^{\alpha}\!_{\mu}\!^{\beta}\!_{\rho} \nn\\
&+6\pi(1-n) \int_\Sigma d^{d-2}x \sqrt{\gamma} \Big( \mathcal R^{\alpha\mu\beta\rho}  \mathcal R_{\mu\nu\rho\sigma} \big[ n^\nu_i n^{\sigma i} n_{\alpha j} n_\beta^{j}-n^\nu_i n_{\beta}^i n _{\alpha j} n^{\sigma j} \big] 
-\frac12 K_2^2 +\frac12 K_4 \nn\\
&+\frac{1}{12} K K_{\mu\nu} \mathcal R^{\mu\nu} +\frac{7}{12}K_{\mu\alpha} \mathcal R^{\alpha \nu} K_{\nu}^{\mu} - \frac{1}{12} \mathcal R K_2- \frac43 K^{\mu\alpha} K^{\nu\beta} \mathcal R_{\mu\nu\alpha\beta}\Big)\,,
\end{align}
\begin{align}
&\int_{\tilde{\mathcal M}_n} d^dx \sqrt{g} RR_{\mu\nu\rho\sigma} R^{\mu\nu\rho\sigma}= n\int_{\mathcal M} d^dx \sqrt{g} R  R_{\mu\nu\rho\sigma}  R^{\mu\nu\rho\sigma}
\nn\\
&+ 4\pi (1-n)\int_\Sigma d^{d-2}x \sqrt{\gamma}  \Big(\big[  R_{\mu\nu\rho\sigma}  R^{\mu\nu\rho\sigma} +2 R  R^{\mu\nu\rho\sigma} n_{\mu i} n_\rho^i n_{\nu j} n_\sigma^j \big] 
-K_2K^2+ K_2^2 -2 \mathcal R K_2\Big)\,, 
\end{align}
\begin{align}
&\int_{\tilde{\mathcal M}_n} d^dx \sqrt{g} R_\alpha\!^\sigma R_{\mu\nu\rho\sigma} R^{\mu\nu\rho\alpha}= n\int_{\mathcal M}  d^dx \sqrt{g} R_\alpha\!^\sigma  R_{\mu\nu\rho\sigma}  R^{\mu\nu\rho\alpha}\nn \\
&+2 \pi (1-n)\int_\Sigma d^{d-2}x \sqrt{\gamma}  \Big( 2 R^{\mu\nu\rho\alpha}  R_\alpha\!^\sigma \big[ n_{\mu i} n_\rho^i n_{\nu j} n_\sigma^j-n_{\mu i} n_\sigma^i n_{\nu j} n_\rho^j]+  R_{\mu\nu\rho\sigma}  R^{\mu\nu\rho\alpha} n_{\alpha i}n^{\sigma i}
\nn \\
& - K_2^2+K_4 -\frac{1}{3} K K_{\mu\nu} \mathcal R^{\mu\nu} -\frac{19}{3}K_{\mu\alpha} \mathcal R^{\alpha \nu} K_{\nu}^{\mu} + \frac{1}{3} \mathcal R K_2- \frac{8}{3} K^{\mu\alpha} K^{\nu\beta} \mathcal R_{\mu\nu\alpha\beta}\Big)\,,
\end{align}
\begin{align}
&\int_{\tilde{\mathcal M}_n} d^dx \sqrt{g} R^{\nu\sigma} R^{\mu\rho} R_{\mu\nu\rho\sigma}= n\int_{\mathcal M} d^dx \sqrt{g} R^{\nu\sigma}  R^{\mu\rho}  R_{\mu\nu\rho\sigma}\nn\\
&+2\pi (1-n) \int_\Sigma d^{d-2}x \sqrt{\gamma} \Big( 2 R_{\mu\nu\rho\sigma}  R^{\mu\rho} n_i^\nu n^{\sigma i}+ R^{\mu\rho}  R^{\nu\sigma} \big[ n_{\mu i} n_\rho^i n_{\nu j} n_\sigma^j-n_{\mu i} n_\sigma^i n_{\nu j} n_\rho^j]\nn\\
& -K_2K^2 +  K_2^2-\frac{23}{12} K K_{\mu\nu} \mathcal R^{\mu\nu} +\frac{7}{12}K_{\mu\alpha} \mathcal R^{\alpha \nu} K_{\nu}^{\mu} - \frac{1}{12} \mathcal R K_2+ \frac{2}{3} K^{\mu\alpha} K^{\nu\beta} \mathcal R_{\mu\nu\alpha\beta}\Big)\,.
\end{align}
\pagebreak
\section{Table of Integrals}
\begin{table}[ht]
\begin{center}
\centering
\scalebox{0.9}{
\setlength{\tabcolsep}{0.0cm}
\renewcommand{\arraystretch}{1.0}
\begin{tabular}{|c|c|c|c|c|c|c|c|c|c|c|c|c|}
\hline 
$ \displaystyle \int_\Sigma d^{d-2}x \sqrt{\gamma}\cdots$ &$AdS^C_5$&$AdS^{S^2}_5$&$AdS^{S^3}_6$&$AdS^{S^4}_7$&$AdS^{S^5}_8$&$AdS^{S^6}_9$ &$AdS^C_5$&$AdS^{S^2}_5$&$AdS^{S^3}_6$&$AdS^{S^4}_7$&$AdS^{S^5}_8$&$AdS^{S^6}_9$\\
 \hline
$  K^4 $& $2 $& $64 $& $162 $&$\frac{2048}{3} $ &$625 $ & $\frac{6912}{5}$& 0&0&0&0&0 &0\\
\hline
$  K_2 K^2$ &2 &32 &54 &$\frac{512}{3}$ &125&$\frac{1152}{5}$ & 0 &0&0&0&0&0\\
\hline
$  KK_3=K_2^2 $ & 2&16 & 18& $\frac{128}{3}$& 25&$\frac{192}{5}$& 0 &0&0&0&0&0\\
\hline
$ K_4 $ & 2&8 &6 & $\frac{32}{3}$&5&$\frac{32}{5}$& 0&0&0&0&0&0\\
\hline
$ \mathcal R K^2     $& 0&0&0&0&0&0& $-40$&$-320$&$-540$&$-1792$&$-1400$&$-\frac{13824}{5}$ \\
\hline
$ K K_{\mu\nu} \mathcal R^{\mu\nu}  $   &0 &0&0&0 &0&0& $-8$&$-64$&$-90$&$-256$&$-175$&$-\frac{1536}{5}$ \\
\hline
$ K_{\mu\alpha}\mathcal R^{\alpha\nu} K_{\nu}\!^\mu $     &0 & 0&0&0&0&0& $-8$&$-32$&$-30 $&$-64$&$-35$&$-\frac{256}{5}$ \\
\hline
$ \mathcal R K_2   $& 0&0&0&0&0&0&$-40$& $-160$&$-180$&$-448$&$-280$&$-\frac{2304}{5}$\\
\hline
$ K^{\mu\nu} K^{\alpha\beta} \mathcal R_{\mu\alpha\nu\beta} $   & 0&0 &0 &0 &0&0&0& $-8$&$-12$&$-32$&$-20$&$-32$ \\
\hline
\hline 
$ \displaystyle \int_{\tilde{\mathcal{M}}} d^dx\sqrt{g} \cdots$ &$AdS^C_5$&$AdS^{S^2}_5$&$AdS^{S^3}_6$&$AdS^{S^4}_7$&$AdS^{S^5}_8$&$AdS^{S^6}_9$ &$AdS^C_5$&$AdS^{S^2}_5$&$AdS^{S^3}_6$&$AdS^{S^4}_7$&$AdS^{S^5}_8$&$AdS^{S^6}_9$\\
\hline
$ RR_2 $    
&0&32&108&512&500&1152&$-80$&$-640$&$-1080$&$-3584$&$-2800$&$-\frac{27648}{5}$\\
\hline
$  R^3 $  
&0 & $0$& $0$& $0$&$0$&0&0&0&0&0&0&0 
\\
\hline$  R_3 $  
&0 & $0$& $0$& $0$&$0$&0&$-48$&$-384$&$-540$&$-1536$&$-1050$&$-\frac{9216}{5}$\\
\hline
$ R^{\mu\nu\rho\sigma} R_{\nu\alpha\sigma\beta} R^{\alpha}\!_{\mu}\!^{\beta}\!_{\rho} $
&0&24&36&96&60&96&12&0&$-36$&$-128$&$-90$&$-\frac{768}{5}$\\     
\hline
$  RR_{\mu\nu\rho\sigma}R^{\mu\nu\rho\sigma} $
&0 & 64& 144& 512&400&768&$-320$&$-1280$&$-1440$&$-3584$&$-2240$ &$-\frac{18432}{5}$\\
\hline
$ R_{\mu\nu\rho\sigma}R^{\mu\nu\rho\alpha} R_{\alpha}\!^{\sigma} $   
&0 & 16& 24& 64& 40&64&$-80$&$-384$&$-384$&$-\frac{2560}{3}$&$-480$ &$-\frac{3584}{5}$\\
\hline
$  R_{\mu\nu\rho\sigma} R^{\mu\rho} R^{\nu\sigma} $  
&0& 32& 72& 256&200&384&$-28$&$-224$&$-324$&$-\frac{2816}{3}$&$-650$&$-1152$\\
\hline
$\times$& $\frac{\pi H z \f^{1/2}}{L a^3} $& $ \frac{\pi z \f^{1/2}}{L a^2}$& $ \frac{\pi^2}{a}$&$\frac{\pi^2 L}{z\f^{1/2}}$ &$ \frac{\pi^3a L^2}{z^2\f}$&$\frac{\pi^3 a^2 L^3}{ z^3\f^{3/2}}$&$\frac{\pi H \f^{1/2} }{L a z}$&$\frac{\pi \f^{1/2} }{Lz }$&$\frac{{\pi^2}a}{z^2}$&$\frac{\pi^2 a^2 L}{z^3 \f^{1/2}}$&$\frac{\pi^3 a^3 L^2}{z^4 \f}$ &$\frac{\pi^3 a^4 L^3}{z^5\f^{3/2} }$\\
\hline
\end{tabular}}
\end{center} 
\caption{The value of each integral can be found by multiplying its value to a proper coefficient shown on the row by $\times$ sign. For example: $\int_{S^2} d^{2}x \sqrt{\gamma} K^4 = 64\pi z\f^{1/2}/L a^2$.
For the second part of the table each value additionally must be multiply by $\pi (n-1)$.}
\end{table}

\end{document}